\documentclass[12pt]{article}
\usepackage{amssymb}

\newcommand{\pd}{\partial}
\newcommand{\D}{\mathrm{d}}
\newcommand{\e}{\mathrm{e}}
\newcommand{\ageq}{\stackrel{>}{\scriptstyle\sim}}
\newcommand{\aleq}{\stackrel{<}{\scriptstyle\sim}}
\newcommand{\OO}{{\cal O}}
\newtheorem{claim}{Claim}[section]
\newtheorem{proposition}[claim]{Proposition}

\begin{document}
\title{Large gaps in point-coupled periodic \\ systems of manifolds}
\date{}
\author{Jochen Br\"uning$^{a}\!$, Pavel Exner$^{b,c}\!$, Vladimir A. Geyler$^{a,d}\!$}
\maketitle

\begin{quote}
{\small {\em a) Institut f\"ur Mathematik, Humboldt Universit\"at zu Berlin,
\\ \phantom{a) } Rudower Chaussee 25, 12489 Berlin, Germany
 \\ b) Nuclear Physics Institute, Academy of Sciences,
25068 \v Re\v z \\ \phantom{a) }near Prague, Czechia
 \\ c) Doppler Institute, Czech Technical
University, B\v rehov{\'a} 7,\\ \phantom{a) }11519 Prague,
Czechia}
 \\ {\em d) Department of Mathematical Analysis, Mordovian State \\
 \phantom{a) } University,  430000 Saransk, Russia;}\\
 \phantom{a) }\texttt{bruening@mathematik.hu-berlin.de}, \texttt{exner@ujf.cas.cz},
 \\ \texttt{geyler@mrsu.ru} }
\end{quote}

\begin{abstract}
\noindent We study a free quantum motion on periodically
structured manifolds composed of elementary two-dimensional
``cells'' connected either by linear segments or through points
where the two cells touch. The general theory is illustrated with
numerous examples in which the elementary components are spherical
surfaces arranged into chains in a straight or zigzag way, or
two-dimensional square-lattice ``carpets''. We show that the spectra
of such systems have an infinite number of gaps and that the latter
dominate the spectrum at high energies.
\end{abstract}


\setcounter{equation}{0}
\section{Introduction}

The spectral behaviour of periodic systems is of a great
importance. Having typically a band structure, such spectra differ
by the number and structure of the gaps. For usual Schr\"odinger
operators the number of gaps is generically infinite in the
one-dimensional situation and finite in higher dimensions.
Moreover, the gap widths decrease as the energy tends to infinity,
the rate of  decay being tied to the regularity of the
potential.

In case of a singular periodic interaction the gaps may not close.
A canonical example is the Kronig-Penney (KP) model, i.e. a chain of
$\delta$-potentials where the gaps are asymptotically constant
\cite{AGHH}. Even more singular couplings like generalized point
interactions may exhibit gaps which are growing at the same rate
as the bands \cite{EGr}, or even grow while the band widths are
asymptotically constant. A typical example of such a behaviour is
a modification of the KP model with a chain of the so-called
$\delta'$-interactions \cite{AGHH}. This behaviour is not
restricted to one dimension; similar results can be derived e.g.
for lattice graphs with appropriate boundary conditions coupling
the wave functions at the vertices \cite{Ex, EGa}.

Large gaps has interesting physical consequences. For instance,
the corresponding Wannier-Stark problem in which we add a linear
background potential to a periodic chain of $\delta'$-interactions
has counterintuitive properties: the absolutely continuous
spectrum of the corresponding Hamiltonian is empty \cite{AEL}, and
in fact, the spectrum is known to be pure point for ``most
values'' of the potential slope \cite{ADE}.
These results can be explained by observing
that tilted gaps represent classically forbidden
regions and that their large widths prevent the particle of propagating
over long distances.

On the other hand, the physical meaning of the 
$\delta'$-coupling remained unclear for a long time. Recently it 
has been demonstrated that this interaction can be approximated 
in the norm resolvent sense by a family of Schr\"odinger 
operators -- see \cite{ENZ} where also a bibliography to the 
problem is given -- but previous studies brought some 
interesting non-potential approximations. An interesting example 
is given by a ``bubble scattering'' in which two halflines are 
attached to the surface of a sphere -- see \cite{Ki} and also 
\cite{ETV, Br}. Such a system typically exhibits numerous 
resonances but the background transmission probability dominates 
and vanishes at the limit of large energies. This observation is 
of importance because systems of a mixed dimensionality are not 
just a mathematicians toy, but they can model real objects such 
as a fullerene  molecule coupled to a pair of nanotubes 
\cite{Ka}.\footnote{Another model for such systems could be that 
of manifolds connected smoothly by thin tubes. Existence of gaps 
in this setting was demonstrated recently by Post \cite{Po}.} 

The aim of the present paper is to study systems with components
of different dimension in the periodic setting. We intend to
demonstrate that the structure of the configuration space
in this case is reflected in the gap behaviour. After describing a
general method to couple periodic systems of spheres,
either joining them by line segments or directly through points
where they touch, we will discuss in Sections~3-5 a number of
examples. The results, summarized in Proposition~\ref{final}, show
that in all the considered cases the number of gaps are infinite
and the gap-to-band width ratio increases with the band index.
The estimated growth is slower than in the case of the
$\delta'$-interaction, and it is slower for a two-dimensional lattice than
for a linear chain, but it is still powerlike for spheres joined
by linear segments, thus confirming our conjecture that the effect
is related to the change in dimensionality the particle must
undergo. Even for a tighter coupling, however, where the spheres
are coupled directly through contact points, the gap-to-band ratio
is still logarithmically increasing.


\setcounter{equation}{0}
\section{General theory}
\subsection{Building blocks of the Hamiltonian} \label{Hamil}

Suppose that $X_0$ is a two-dimensional Riemann manifold. By $H_0$
we denote a Schr\"odinger operator,
 $$ 
 H_0 = |g|^{-1/2} (-i\pd_j-A_j) |g|^{1/2} g^{jk} (-i\pd_k-A_k) +
 V\,,
 $$ 
on $L^2(X_0, |g|^{1/2} \D x)$ with smooth vector and scalar
potentials. The formalism we are going to describe extends easily
to the case $\dim X_0= 3$ but we will limit
ourselves here to referring to \cite{BG2} for guidelines concerning such
a generalization. The metric structure of $X_0$ is fixed and we
will employ the shorthand notation $L^2(X_0)$ for simplicity in
the following. Let further $X_j,\: j=1,\dots,n$, be a finite or
semiinfinite line segment which can be identified with the
interval $[0,d_j),\: 0<d_j<\infty$. No external potentials are
supposed to act on the particle on $X_j$, i.e. we consider the
free operators $H_j = -\D^2/\D x^2$ on $L^2(X_j)$ with Neumann's
condition at the endpoints (the ``right'' endpoint $x=d_j$
requires a boundary condition only if $d_j<\infty$) as the
building blocks of the system Hamiltonian.

As we have said above we consider systems with configuration
space consisting of infinite number of copies of a manifold which
are connected either by isolated points common for the pair of
neighbouring copies, or by line segments connecting such points.
We will concentrate on the latter case which is more complicated.
The former one can be regarded as the limiting situation where the
length of the connecting segments tends to zero, and the
corresponding modification of the formalism is easy.

A building block of our model is thus a ``hedgehog manifold''
obtained by attaching each segment $X_j$ to $X_0$ at a point
$q_j\in X_0$, or more exactly, by identifying the point $0\in X_j$
with $q_j\in X_0$; we suppose that all the connection points $q_j$
are mutually different. The topological space constructed in this
way will be denoted as $\hat X$; it can be endowed with a natural
measure which restricts to the Riemannian measure on $X_0$ and to the
Lebesgue measure on each $X_j\,,\; j=1,\dots,n$. This yields the
identification
 $$ 
 L^2(\hat X)= L^2(X_0)\oplus L^2(X_1)\oplus \cdots L^2(X_n)
 $$ 
for the Hilbert state space of the system.

By $S_0$ we denote the restriction of the operator $H_0$ defined
above to the family of functions
 $$ 
 \left\{\, f\in{\mathcal{D}}(H_0)\,:\: f(q_1)= \cdots= f(q_n)=0
 \, \right\}\,,
 $$ 
which obviously makes sense as long as $\dim X_0\le 3$. In a
similar way we use the symbol $S_j,\, j=1,\dots,n\,$, for the
restriction of $H_j$ to the set $\left\{ f\in{\mathcal{D}}
(H_j):\, f(0)=0 \right\}$. The Schr\"odinger operators we consider
are {\em by definition} self-adjoint
extensions of the symmetric operator $S= S_0\oplus S_1\oplus
\cdots \oplus S_n$; their construction is a standard
matter discussed in numerous papers starting with \cite{ES1}. The
most efficient way to describe them is based on a bijective
correspondence with the Lagrangian planes in $\mathcal{G}\times\mathcal{G}$,
where $\mathcal{G}= \mathbb{C}^{2n}$. To describe
it, we introduce the boundary-value operators
 $$ 
 \Gamma^1,\, \Gamma^2:\; {\mathcal{D}}(S^*) \to {\mathcal{G}}\,,
 $$ 
by
 \begin{eqnarray}
 \Gamma^1(f) &\!:=\!& \left( a(f_0,q_1), \dots, a(f_0,q_n),
 -f'_1(0), \dots, -f'_n(0) \right)\,, \nonumber \\
 \Gamma^2(f) &\!:=\!& \left( b(f_0,q_1), \dots, b(f_0,q_n),
 f_1(0), \dots, f_n(0) \right)\,. \label{Gamma}
 \end{eqnarray}
Here $a(f_0,q_j)=: a_j(f_0)$ and $b(f_0,q_j)=: b_j(f_0)$ are the
leading-term coefficients of the asymptotics of $f_0$ in the
vicinity of the point $q_j$ as determined in \cite{BG1, BG2}, or the
generalized boundary values $2\pi(\dim X_0\!-1) L_0$ and $L_1$,
respectively, in the terminology of \cite{ES1, ES2}.

Let $\Lambda$ be a Lagrangian plane in $\mathcal{G}\times
\mathcal{G}$, i.e. $\Lambda^\perp= \Lambda$ with
respect to the skew-Hermitean product $[x|y]:= \langle x_1|y_2
\rangle - \langle x_2|y_1 \rangle$ in $\mathcal{G}\times
\mathcal{G}$. Then any restriction of the adjoint operator $S^*$ to
a family of functions from ${\mathcal{D}}(S^*)$ specified by the
boundary condition $(\Gamma^1 f, \Gamma^2 f)\in\Lambda$ is a
self-adjoint operator which we denote $H^\Lambda$. Recall that
a Lagrangian plane is, in general, the graph of a self-adjoint
operator $L:\, {\mathcal{G}} \to {\mathcal{G}}$ so that the above
boundary condition can be rewritten as $\Gamma^2 f= L(\Gamma^1 f)$.
To avoid problems with the invertibility of $L$ one can view
$\Lambda$ also as the graph of a ``multivalued'' operator in
$\mathcal{G}$, in other words, one may describe it through a
relation $Lx=My$, $(x,y)\in \mathcal{G}\times \mathcal{G}$,
where $L,M:\: {\mathcal{G}}\to {\mathcal{G}}$ are linear operators
satisfying the conditions \cite{KS1}:
 \begin{description}
 \item{\em (i)} $\;LM^*= ML^*$,
 \vspace{-1.8ex}
 \item{\em (ii)} $\; \mathrm{rank}(L,-M)=n$\,.
 \end{description}


\subsection{The resolvent}

We are concerned with spectral properties of
the said self-adjoint extensions, which are as usual defined from
the resolvent. The latter is expressed here by Krein's formula
\cite[App.~A]{AGHH}: If we denote by $H^0$ the decoupled operator
$H_0\oplus H_1\oplus \cdots H_n$, then we have
 \begin{equation} \label{krein}
 (H^\Lambda -z)^{-1} = (H^0 -z)^{-1} -\gamma(z)
 [Q(z)-\Lambda]^{-1} \gamma^*(z)
 \end{equation}
for any $z$ in the resolvent set, in particular for $z\notin \mathbb{C}
\setminus \mathbb{R}$, where the operator $\gamma(z):
{\mathcal{G}} \to {\mathcal{H}}$ is given by the formula
 $$ 
 \gamma(z) := \left( \Gamma^1\upharpoonright {\mathcal{N}}_z
 \right)^{-1}, \quad {\mathcal{N}}_z = \mathrm{Ker}\,
 (S^*-\!z)\,,
 $$ 
and $Q(z):\: {\mathcal{G}}\to {\mathcal{G}}$ is defined as
 $$ 
 Q(z):= \Gamma^2 \gamma(z)\,.
 $$ 
Then the inverse $[Q(z)-\Lambda]^{-1}$ exists for all non-real $z$. To
find an explicit expression for the Green function of the operator
$H^\Lambda$ from (\ref{krein}) we need to know the Green function
$G_0$ of $H^0$.

Notice first that it is easy to find the Green function $G_j$ of
$H_j$: one has
 \begin{equation} \label{segmentG}
 G_j(x,x';z) = {{\cosh\left[-\sqrt{-z}(d_j-|x-x'|)\right]
 + \cosh\left[-\sqrt{-z}(d_j-(x+x')\right]} \over 2\sqrt{-z}\,
 \sinh\left[-\sqrt{-z}d_j \right]} \,.
 \end{equation}
Using the natural decomposition ${\mathcal{G}}= \mathbb{C}^{2n} =
\mathbb{C}^n \times \mathbb{C}^n$ we write the matrix
representation of the operator $[Q(z)-\Lambda]^{-1}$ in block
form,
 \begin{equation} \label{block}
 [Q(z)-\Lambda]^{-1} = \left\lbrack \begin{array}{cc} T(z) & W(z) \\
 U(z) & V(z) \end{array} \right\rbrack\,.
 \end{equation}

Since ${\mathcal{H}}={\mathcal{H}}_0\oplus {\mathcal{H}}_1\oplus \cdots
{\mathcal{H}}_n$, the Green function of the operator $H^\Lambda$ can be
represented as a matrix of integral kernels  of
operators acting from ${\mathcal{H}}_k$ to ${\mathcal{H}}_j$,
 \begin{equation} \label{matrix form}
 G^\Lambda(x,x';z)= \left( G_{jk}^\Lambda(x_j,x'_k;z) \right)_{0\le
 j,k\le n} \quad \mathrm{with} \quad x_j\in X_j,\: x'_k\in X_k\,.
 \end{equation}
Let $(\xi_1,\dots,\xi_n, \eta_1,\dots,\eta_n) \in {\mathcal{G}} =
\mathbb{C}^n \times \mathbb{C}^n$, then a direct calculation
yields
 \begin{eqnarray*} 
 \lefteqn{ \gamma(z)(\xi_1,\dots,\xi_n, \eta_1,\dots,\eta_n)}
 \\ && = \left(
 \sum_{j=1}^n G_0(\cdot,q_j;z)\xi_j,\, G_1(\cdot,0;z)\eta_1,
 \dots, G_n(\cdot,0;z)\eta_n \right)\,.
 \end{eqnarray*}
This implies the adjoint operator action,
 $$ 
 \gamma^*(\bar z)(f_0,f_1,\dots,f_n) =
 (\xi_1,\dots,\xi_n, \eta_1,\dots,\eta_n)\,,
 $$ 
where
 $$ 
 \xi_j= \int_{X_0} G_0(q_j,x;z) f_0(x)\,|g(x)|^{1/2}\, \D x\,,
 \quad \eta_j = \int_{X_j} G_j(0,x;z) f_j(x)\, \D x\,.
 $$ 
The matrix $Q(z)$ then has block-diagonal form
 $$ 
 Q(z) = \left\lbrack \begin{array}{cc} Q^{11}(z) & 0 \\
 0 & Q^{22}(z) \end{array} \right\rbrack\,,
 $$ 
here $Q^{11}(z)$ coincides with the $Q$-matrix $Q_0$ for the pair
$(S_0,H_0)$. Recall that the $Q$-function in the Krein
formula always corresponds to a pair of a self-adjoint operators
and fixed symmetric restriction. In the present case it has
the form
 \begin{equation} \label{Qmanif}
 Q_0^{jk}(z) = G_0^\mathrm{ren}(q_j,q_k;z)\,,
 \end{equation}
where $G_0^\mathrm{ren}$ is the renormalized Green's function
obtained from $G_0$ by subtracting the diagonal singularity,
 $$ 
 G_0^\mathrm{ren}(x_0,x'_0;z) =
 \cases{
 G_0(x_0,x'_0;z)\,, &if
 $x_0\ne x'_0$;\cr
 \displaystyle\lim_{y_0\to x_0} \left\lbrack G_0(x_0,y_0;z) + {1\over 2\pi}
 \ln \rho(x_0,y_0) \right\rbrack\,, &if
 $x_0=x'_0$.\cr}
 $$ 
Here $\rho(x_0,y_0)$ denotes the geodesic distance on $X_0$. On
the other hand, the matrix $Q^{22}(z)$ is diagonal,
 $$ 
 Q^{22}_{jk}(z) = \delta_{jk} G_j(0,0;z)\,.
 $$ 
Using the above formulae we can write the matrix element kernels
in (\ref{matrix form}) more explicitly,
 $$ 
 G_{jk}(x_j,x'_k;z) = \delta_{jk} G_j(x_j,x'_j;z) -
 K_{jk}(x_j,x'_k;z)\,,
 $$ 
where
 \begin{eqnarray*} 
 K_{00}(x_0,x'_0;z) &\!=\!& \sum_{j,k=1}^{n} t_{jk}(z)
 G_0(x_0,q_j;z) G_0(q_k,x'_0;z) \,, \\
 K_{0k}(x_0,x'_k;z) &\!=\!& G_k(0,x'_k;z) \sum_{j=1}^n w_{jk}(z)
 G_0(x_0,q_j;z)\,, \quad k>0\,, \\
 K_{j0}(x_j,x'_0;z) &\!=\!& G_j(x_j,0,;z) \sum_{k=1}^n u_{jk}(z)
 G_0(q_k,x'_0;z)\,, \quad j>0\,, \\
 K_{jk}(x_j,x'_k;z) &\!=\!& v_{jk}(z) G_j(x_j,0,;z)
 G_k(0,x'_k;z)\,, \quad j,k>0\,;
 \end{eqnarray*}
the coefficients refering to the block representation (\ref{block}), $\left(
t_{jk}(z) \right)= T(z)$, etc., can be in principle computed explicitly.


\subsection{Coupling hedgehog manifolds}

In the next step we are going to glue together the building blocks
considered so far. To begin with, we consider such a manifold
$\hat X$ and select some number of finite segments of
lengths $d_1,\dots,d_s,\: 1\le s\le n$. At the same time, we fix a
finite number of distinct points $p_1,\dots,p_m\in X_0$ such that
$\{p_1,\ldots,p_n\}\cap\{q_1,\ldots,q_n\}=\emptyset$.
We fix a Hamiltonian $H^\Lambda$ on $\hat X$ and consider its restriction
$\tilde S$ to the set of functions
 $$ 
 \left\{\, f\in{\mathcal{D}}(H^\Lambda)\,:\: f(p_1)= \cdots=
 f(p_m) =f(d_1) =\cdots= f(d_s)=0 \, \right\}.
 $$ 
Let us find the $\cal Q$-matrix of the pair $(\tilde
S,H^\Lambda)$ which is a $(m+s)\times(m+s)$ matrix $\tilde Q(z)$ with
block structure,
 $$ 
 \tilde Q(z) = \left\lbrack \begin{array}{cc} \tilde Q^{11}(z)
 & \tilde Q^{12}(z) \\ \tilde Q^{21}(z) & \tilde Q^{22}(z)
 \end{array} \right\rbrack\,.
 $$ 
Using the formula for the Green function of $H^\Lambda$ we can
write the elements of the above matrix as
 \begin{eqnarray}
 \tilde Q_{jk}^{11}(z) &\!=\!& \delta_{jk}
 G_0^\mathrm{ren} (p_j,p_j;z) + (1-\delta_{jk}) G_0 (p_j,p_k;z)
 \nonumber
 \\ && - K_{00}(p_j,p_k;z)\,,\qquad 1\le j,k\le m\,, \nonumber \\
 \tilde Q_{jk}^{12}(z) &\!=\!& -K_{0k}(p_j,d_k;z)\,,\qquad 1\le
 j\le m\,,\, 1\le k\le s\,, \label{tildeQ} \\
 \tilde Q_{jk}^{21}(z) &\!=\!& -K_{j0}(d_j,p_k;z)\,,\qquad
 1\le j\le s\,,\, 1\le k\le m\,, \nonumber \\
 \tilde Q_{jk}^{22}(z) &\!=\!& \delta_{jk} G_j (d_j,d_j;z)
 -K_{jk}(d_j,d_k;z)\,,\quad 1\le j,k\le s\,. \nonumber
 \end{eqnarray}
Recall that $G_0^\mathrm{ren}$ denotes the renormalized Green's
function; we drop of course the superscript whenever the two
arguments are different.

The coupling will be realized through conditions relating
the generalized boundary values. We will not strive for at most
generality, however, because formulae encompassing manifolds with
arbitrary $n,m$ would be rather cumbersome. We
will instead discuss in some detail properties of a
quantum particle living on chained manifolds of different
dimensions, i.e. the case $m=n=1$; later on we will extend the argument
to a particular situation with $m=n=2$.

Consider, therefore, a manifold $X_0$ on which a pair of mutually
different points $p,q$ is selected. At $q$, a segment of a length
$d$ is attached, while $p$ is a ``socket'' to which another
``tailed'' manifold can be coupled. In analogy with (\ref{Qmanif})
we introduce the matrix
 \begin{equation} \label{Q_0}
 Q_0(z) = \left\lbrack \begin{array}{cc} G_0^\mathrm{ren} (q,q;z)
 & G_0(p,q;z) \\ G_0(q,p;z) & G_0^\mathrm{ren} (p,p;z)
 \end{array} \right\rbrack\,,
 \end{equation}
and similarly, the segment will be characterized by
 \begin{equation} \label{Q_1}
 Q_1(z) = \left\lbrack \begin{array}{cc} G_1 (0,0;z)
 & G_1(0,d;z) \\ G_1(d,0;z) & G_1 (d,d;z)
 \end{array} \right\rbrack\,.
 \end{equation}
Using (\ref{segmentG}) we find
 $$ 
 Q_1^{jk}(z) = {\delta_{jk}\over\sqrt{-z}} \coth\left(
 \sqrt{-z}d \right) + {{1- \delta_{jk}}\over \sqrt{-z} \sinh\left(
 \sqrt{-z}d \right)}\,,
 $$ 
or
 \begin{equation} \label{explQ_1}
 Q_1^{jk}(z) = {\delta_{jk}\over k} \cot(kd) + {{1- \delta_{jk}}
 \over k \sin(kd)}
 \end{equation}
in the usual momentum notation, $k:= i\sqrt{-z}$ for $z\in
\mathbb{C} \setminus \mathbb{R}_+$.

The operator $H^\Lambda$ on $\hat X$ is specified by the
boundary conditions at the point $q$ identified with the left
endpoint of the segment, $0\in[0,d)$. In general, these conditions can be
given in the form,
 \begin{eqnarray}
 b(f_0,q) &\!=\!& \alpha f'_1(0) + \beta a(f_0,q)\,, \nonumber \\
 f_1(0) &\!=\!& \gamma f'_1(0) -\bar\alpha a(f_0,q) \label{1bc}\,,
 \end{eqnarray}
with $\beta,\, \gamma\in\mathbb{R}$ and $\alpha\in\mathbb{C}$; we
suppose $\alpha\ne 0$ such that the manifold $X_0$ and the segment
are coupled in a nontrivial way. For the sake of simplicity we
will restrict ourselves to the case where $\beta=\gamma=0$, i.e.
 \begin{equation} \label{minbc}
 b(f_0,q) = \alpha f'_1(0)\,, \quad f_1(0) = -\bar\alpha
 a(f_0,q)\,.
 \end{equation}
This can be regarded as a ``minimal'' coupling between the two
configuration-space components, because in the ``switched-off
state'', $\alpha=0$, the manifold Hamiltonian contains no point
interaction at the point $q$ and the segment part satisfies the
Dirichlet condition at $x_1=0$. Notice, however, that there are
other natural choices such as
 $$ 
 \alpha= \sqrt{2\rho\over\pi}\,,\quad \beta=
 -\pi(1+\ln\sqrt\rho)\,, \quad \gamma= 2\rho\,,
 $$ 
which describes the particle passing through the
junction at a low energy if the segment models a thin tube of
radius $\rho$ -- cf.~\cite{ES2}.

The boundary condition (\ref{minbc}) can be cast into the form
given in Sec.~\ref{Hamil} if we choose $M$ as the $2\times2$ unit
matrix and
\begin{equation} \label{B1}
 L := \left\lbrack \begin{array}{cc} 0
 & \alpha \\ \bar\alpha & 0
 \end{array} \right\rbrack\,.
\end{equation}
The $Q$-matrix entering Krein's formula for the operator
$H^\Lambda$ can be expressed in terms of the matrices (\ref{Q_0}) and
(\ref{Q_1}) as
 \begin{equation} \label{B2}
 Q(z) = \left\lbrack \begin{array}{cc} Q_0^{11}
 & 0 \\ 0 & Q_1^{11} \end{array} \right\rbrack\,.
 \end{equation}
From (\ref{B1}) and (\ref{B2}) we find
 $$ 
 [Q(z)-\Lambda]^{-1} = {1\over Q_0^{11}(z) Q_1^{11}(z) -|\alpha|^2}
 \left\lbrack \begin{array}{cc} Q_1^{11}(z)
 & -\alpha \\ -\bar\alpha & Q_0^{11}(z) \end{array}
 \right\rbrack\,,
 $$ 
and therefore
 \begin{eqnarray*} 
 G_{00}^\Lambda(x_0,x'_0;z) &\!=\!& G_0 (x_0,x'_0;z)
 - {Q_1^{11}(z) \over Q_0^{11}(z) Q_1^{11}(z) \!-\!|\alpha|^2}\,
 G_0 (x_0,q;z) G_0 (q,x'_0;z) , \\
 G_{01}^\Lambda(x_0,x'_1;z) &\!=\!&
 {\alpha \over Q_0^{11}(z) Q_1^{11}(z) \!-\!|\alpha|^2}\,
 G_0 (x_0,q;z) G_1 (0,x'_1;z) \,, \\
 G_{10}^\Lambda(x_1,x'_0;z) &\!=\!&
 {\bar\alpha \over Q_0^{11}(z) Q_1^{11}(z) \!-\!|\alpha|^2}\,
 G_1(x_1,0;z) G_0 (q,x'_0;z) \,, \\
 G_{11}^\Lambda(x_1,x'_1;z) &\!=\!& G_1 (x_1,x'_1;z)
 - {Q_0^{11}(z) \over Q_0^{11}(z) Q_1^{11}(z) \!-\!|\alpha|^2}\,
 G_1 (x_1,0;z) G_1 (0,x'_1;z) \,.
 \end{eqnarray*}
Thus we can calculate the matrix elements (\ref{tildeQ}) (the
indices $j,k$ are trivial in the present example and we will
drop them):
 \begin{eqnarray} 
 \tilde Q^{11}(z) &\!=\!& \tilde Q_0^{22}(z)
 - {Q_1^{11}(z) Q_0^{12}(z) Q_0^{21}(z) \over
 Q_0^{11}(z) Q_1^{11}(z) \!-\!|\alpha|^2}\,, \nonumber \\
 \tilde Q^{12}(z) &\!=\!&
 {\alpha Q_1^{12}(z) Q_0^{21}(z) \over
 Q_0^{11}(z) Q_1^{11}(z) \!-\!|\alpha|^2}\,, \label{tildeQel} \\
 \tilde Q^{21}(z) &\!=\!&
 {\bar\alpha Q_0^{12}(z) Q_1^{21}(z) \over
 Q_0^{11}(z) Q_1^{11}(z) \!-\!|\alpha|^2}\,, \nonumber \\
 \tilde Q^{22}(z) &\!=\!& \tilde Q_1^{22}(z)
 - {Q_0^{11}(z) Q_1^{12}(z) Q_1^{21}(z) \over
 Q_0^{11}(z) Q_1^{11}(z) \!-\!|\alpha|^2}\,. \nonumber
 \end{eqnarray}
These formulae can be made even more explicit by plugging in (\ref{explQ_1}).


\subsection{Point-coupled manifolds}

In the same way one can treat the limiting situation when the lengths
of the
connecting segment shrink to zero. Then only the boundary conditions
have be modified. Consider the simplest
case when $X_0$ and $X_1$ are coupled by identifying the points
$p_j\in X_j,\: j=0,1$. The generalized boundary values
(\ref{Gamma}) are then replaced by
 \begin{eqnarray*} 
 \Gamma^1(f_0,f_1) &\!:=\!& \left( a(f_0,p_0), a(f_1,p_1) \right)\,, \\
 \Gamma^2(f_0,f_1) &\!:=\!& \left( b(f_0,p_0), b(f_1,p_1) \right)\,.
 \end{eqnarray*}
Such a coupling was first discussed in \cite{ES3} in the situation
where $X_0$ and $X_1$ are two planes. The four-parameter set of
all possible self-adjoint extensions was described there and the
result adapts easily to more general manifolds. For the sake of
simplicity, however, we will again restrict our attention to the
``minimal'' coupling given by the conditions
 \begin{equation} \label{minbc point}
 b(f_0,p_0) = \alpha a(f_1,p_1)\,, \quad b(f_1,p_1) =
 \bar\alpha a(f_0,p_0)
 \end{equation}
with a complex parameter $\alpha$, decoupled manifolds
corresponding to $\alpha=0$.


\setcounter{equation}{0}
\section{Infinite necklaces}

\subsection{General periodic case}

As an illustration of how to couple ``hedgehog'' manifolds, we are now going
to analyze now the simplest nontrivial case i.e. when the building
blocks discussed above are chained into an infinite ``necklace''.
To define the Hamiltonian we have to specify the boundary
conditions coupling the outer endpoint of the segment
of the first building block, starting at $q$,
to the point $p$ of the second one. The boundary-value operators $\tilde
\Gamma^1$ and $\tilde \Gamma^2$ for the operator $\tilde S$ are of
the form
 \begin{eqnarray*} 
 \Gamma^1(f_0,f_1) &\!:=\!& \left( a(f_0,p), f'_1(d) \right)\,, \\
 \Gamma^2(f_0,f_1) &\!:=\!& \left( b(f_0,p), f_1(d) \right)\,.
 \end{eqnarray*}
Notice the positive sign of $f'_1(d)$ in comparison with
(\ref{Gamma}) which reflects the  orientation of the segment
$[0,d]$.

Consider now a countable family of identical copies of the
manifold $\hat X$, i.e. $\hat X_M=\hat X$ for all
$m\in\mathbb{Z}$ and set $\hat Z:= \bigcup_{m\in\mathbb{Z}} \hat
X_m$. The state Hilbert space of this necklace is
 $$ 
 L^2(\hat Z) = \bigoplus_{m=-\infty}^\infty L^2(\hat X_m)\,.
 $$ 
Schr\"odinger operators on the necklace will be identified with
self-adjoint extensions of the symmetric operator $\hat S:=
\bigoplus_{m\in\mathbb{Z}} \tilde S_m$, where $\tilde S_m :=
\tilde S$ for any $m\in\mathbb{Z}$. Obviously, the
boundary-value space of $\hat S$ is of the form
 $$ 
 \hat\mathcal{G} = \bigoplus_{m=-\infty}^\infty
 \tilde\mathcal{G}_m \qquad \mathrm{with}\quad \tilde\mathcal{G}_m=
 \mathbb{C}^2 \quad \mathrm{for all}\quad  m
 $$ 
and
 $$ 
 \hat\Gamma^j = \bigoplus_{m=-\infty}^\infty
 \tilde\Gamma^j_m \qquad \mathrm{with}\quad \tilde\Gamma^j_m=
 \tilde\Gamma^j \quad \mathrm{for all}\quad m \quad
 \mathrm{and} \quad  j=1,2\,.
 $$ 
Of course, the operator $\hat S$ has infinite deficiency indices,
and therefore plenty of self-adjoint extensions. We restrict our
attention to those which are local in the sense that exactly the
point $d$ of $\hat X_m$ is coupled with the point $p$ of $\hat
X_{m+1}$. Moreover, we will consider the situation when the
coupling $d$ to $p$ and $q$ to $0$ is {\em minimal} in the sense described
above. Consequently, for an
element $g=\{ g_m\}\in \hat\mathcal{G}$ with $g_m= (f_{0,m}, f_{1,m})$ we impose
boundary conditions analogous to (\ref{minbc}):
 $$ 
 b(f_{0,m},p) = \alpha f'_{1,m-1}(0)\,, \quad f_{1,m}(0)
 = -\bar\alpha a(f_{0,m+1},p)\,;
 $$ 
this can be written concisely as
 \begin{equation} \label{minbc-per}
 \hat\Gamma^2 g= L \hat\Gamma^1 g\,,
 \end{equation}
where $L$ is an operator in $\hat\mathcal{G}$ given by a matrix
 $L= (L_{mn})_{m,n\in\mathbb{Z}}$, where
$L_{mn}=0\,$ if $\,|m\!-\!n|\ne 1\,$ and
 $$ 
 L_{m,m+1} = \left\lbrack \begin{array}{cc} 0
 & \alpha \\ 0 & 0
 \end{array} \right\rbrack\,, \quad
 L_{m+1,m} = \left\lbrack \begin{array}{cc} 0
 & 0 \\ \bar\alpha & 0
 \end{array} \right\rbrack\,.
 $$ 
We then infer that the self-adjoint operator $H^L$
specified by the boundary conditions
(\ref{minbc-per}) has the following resolvent
 $$ 
 (\hat H^L -z)^{-1} = (\hat H^0 -z)^{-1} -\hat\gamma(z)
 [\hat Q(z)-L]^{-1} \hat\gamma^*(z)\,,
 $$ 
where $\hat Q(z) = \{ \delta_{mn} \tilde Q(z)\}$. In this way,
the dispersion relation for $\hat H^L$ can be obtained by
introducing the quasimomentum $\theta\in [0,2\pi)$ and
performing the Fourier transformation of the operator $\hat Q(z)-L$.
Thus result is an operator in the space $L^2((0,2\pi))
\otimes \mathcal{G}$ with kernel
 $$ 
 P(\theta,z) := \sum_{m=-\infty}^\infty \left( \hat
 Q_{m0}(z)-L_{m0} \right) \e^{im\theta}
 = \tilde Q(z) - \left\lbrack \begin{array}{cc} 0
 & \alpha\, \e^{i\theta} \\ \bar\alpha\, \e^{-i\theta} & 0
 \end{array} \right\rbrack\,.
 $$ 
The dispersion relation is of the form $\det P(\theta,z)=0$, or
 $$ 
 \det\,\,\left| \begin{array}{cc} \tilde Q_{11}(z)
 & \tilde Q_{12}(z) -\alpha\, \e^{i\theta} \\
 \tilde Q_{21}(z) -\bar\alpha\, \e^{-i\theta} &
 \tilde Q_{22}(z) \end{array} \right| = 0\,,
 $$ 
which is equivalent to
 \begin{equation} \label{spectral}
 \det \tilde Q(z) -\left( \tilde Q_{12}(z) \bar\alpha\,
 \e^{-i\theta} + \tilde Q_{21}(z) \alpha\, \e^{i\theta} \right)
 -|\alpha|^2 = 0\,.
 \end{equation}
As in similar situations, we have isospectrality with
respect to the coupling-constant phase: put
$\varphi=\arg\alpha$, i.e. $\alpha= |\alpha| \e^{i\varphi}$,
then the last condition can be written as
 $$ 
 \det \tilde Q(z) - |\alpha|\left( \tilde Q_{12}(z) \,
 \e^{-i(\theta+\varphi)} + \tilde Q_{21}(z) \,
 \e^{i(\theta+\varphi)} \right) -|\alpha|^2 = 0\,,
 $$ 
which shows that without loss of generality we may restrict
ourselves to the case $\alpha\ge 0$; this we shall assume in
the following. Using the fact that $\tilde Q_{21}^*(z) = \tilde
Q_{12}(z)$ holds for real $z$, the condition (\ref{spectral})
can be rewritten as
 \begin{equation} \label{spectral2}
 \det \tilde Q(z) -|\alpha|^2 = 2\alpha \left( \mathrm{Re}\,
 \tilde Q_{12}(z) \cos\theta +  \mathrm{Im}\,
 \tilde Q_{12}(z) \sin\theta \right)\,.
 \end{equation}
Hence a necessary condition for $z\in{\rm spec}(\hat H)$ is
 \begin{equation} \label{spectral3}
 {|\det \tilde Q(z) -|\alpha|^2| \over 2\alpha |
 \tilde Q_{12}(z) |} \le 1\,.
 \end{equation}
If $\tilde Q_{12}(z) =\tilde Q_{21}(z)$, which is true in
particular if $H^0$ is a real operator (i.e. commutes with the
complex conjugation), the relation (\ref{spectral2}) simplifies
to
 \begin{equation} \label{spectral4}
 \cos\theta ={{\det \tilde Q(z) -|\alpha|^2} \over 2\alpha
 \tilde Q_{12}(z)}\,,
 \end{equation}
and the condition (\ref{spectral3}) becomes necessary and
sufficient. If $H^0$ is real, the
condition (\ref{spectral4}) can be made more explicit: using
(\ref{tildeQel}) and the fact that $Q_j^{11} =Q_j^{22}$ holds for
$j=1,2$, we find after a short computation
 $$ 
 \cos\theta ={{\det Q_0(z)\det Q_1(z) -2\alpha^2 Q_0^{11}(z)
 Q_1^{11}(z) + \alpha^4} \over 2\alpha^2 Q_0^{12}(z)
 Q_1^{12}(z)}\,.
 $$ 
Furthermore, using (\ref{explQ_1}) we get
 \begin{equation} \label{spectral5}
 \cos\theta ={{\det Q_0(k^2)\sin(kd) -2\alpha^2k\, \cos(kd)\, Q_0^{11}(k^2)
 - \alpha^4 k^2 \sin(kd)} \over 2\alpha^2 k\, Q_0^{12}(k^2)}\,.
 \end{equation}


\subsection{Spherical beads}

Since our aim is to present solvable examples, we study next
 the situation when the elementary
building-block manifold $X_0$ is a two-dimensional sphere
$\mathbb{S}^2$ of a fixed radius $a>0$. We parametrize it
by spherical coordinates,
 \begin{eqnarray*} 
 x &\!=\!& a\, \cos\vartheta\, \cos\varphi\,, \\
 y &\!=\!& a\, \cos\vartheta\, \sin\varphi\,, \\
 z &\!=\!& a\, \sin\vartheta\,,
 \end{eqnarray*}
with $\vartheta\in [-\pi/2, \pi/2],\: \varphi\in [0,2\pi)$. We
will assume that there are no external fields, so the starting
operator for construction of the Hamiltonian is the
Laplace-Beltrami operator $\Delta_\mathrm{LB}$ on $\mathbb{S}^2$.
Its Green's function is an integral operator with the kernel
 \begin{equation} \label{sphere G}
 G_0(x,y;z) =  -\, {1\over 4\cos(\pi t)}\, \mathcal{P}_{-{1\over
 2}+t} \left( -\cos\left(\rho(x,y)\over a\right) \right)\,,
 \end{equation}
where $\mathcal{P}_\lambda$ is the Legendre function, $\rho(x,y)$
is the geodetic distance on the sphere, and
 $$ 
 t\equiv t(z):= {1\over 2}\sqrt{1+ 4a^2 z}\,.
 $$ 
This allows us to express the renormalized Green's function,
i.e. we find
 \begin{equation} \label{ren sphere G}
 Q_0^{jj}(z) = -{1\over 2\pi}\, \left\lbrack \psi\left(
 {1\over 2}+t \right) - {\pi\over 2} \tan(\pi t) - \ln 2a +
 C_E \right\rbrack \,,
 \end{equation}
(see e.g. \cite[Tab.~3.9.2]{BE}), where $C_E$ is
Euler's number and $\psi$ the digamma function. We use again the
conventional notation $z=k^2$ for the energy parameter; if there
is no danger of misunderstanding we will often supposes
the dependence of various quantities on $k$.


\subsection{Loose necklaces}

We shall next consider two particular segment-connected periodic
chains: \\ [.5em]
{\sl Example I:} Suppose that the connecting segments are attached
at antipodal points as sketched in Fig.~1 so that the geodesic
distance of the junctions is $\pi a$.
 \begin{figure}
 \setlength\unitlength{1mm}
 \begin{picture}(85,45)(10,15)
 \thicklines
 \put(55.5,30){\circle{15}}
 \put(80.5,30){\circle{15}}
 \put(105.5,30){\circle{15}}
 \put(40.5,30){\line(1,0){8}}
 \put(62.5,30){\line(1,0){11}}
 \put(87.5,30){\line(1,0){11}}
 \put(112.5,30){\line(1,0){8}}
 \put(48.5,30){\circle*{1.2}}
 \put(62.5,30){\circle*{1.2}}
 \put(73.5,30){\circle*{1.2}}
 \put(87.5,30){\circle*{1.2}}
 \put(98.5,30){\circle*{1.2}}
 \put(112.5,30){\circle*{1.2}}
 \put(65,23){\mbox{$I_{n-1}$}}
 \put(91,23){\mbox{$I_n$}}
 \put(52.5,17){\mbox{$\mathbb{S}^2_{n-1}$}}
 \put(79.5,17){\mbox{$\mathbb{S}^2_n$}}
 \put(102.5,17){\mbox{$\mathbb{S}^2_{n+1}$}}
 \end{picture}
 \caption{A loose straight necklace}
 \end{figure}
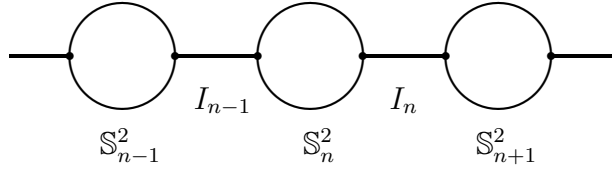
We will denote the segment connecting the spheres $\mathbb{S}^2_n$
and $\mathbb{S}^2_{n+1}$ as $I_n$, with the endpoints
$0^{(n)}\equiv p_1^{(n)} \in \mathbb{S}^2_n$ and $d^{(n)}\equiv
p_3^{(n+1)} \in \mathbb{S}^2_{n+1}$. The lower-index numeration is
somewhat arbitrary and serves just to having a common notation for
the present configuration and that considered below.
 \\ [.5em]
\noindent {\sl Example II:} Alternatively, assume that the
junction points are chosen on one pole and on the equator point, as sketched
in Fig.~2,
 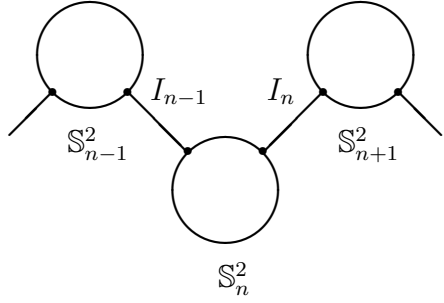
\begin{figure}
 \setlength\unitlength{1mm}
 \begin{picture}(95,45)(10,15)
 \thicklines
 \put(62.5,48){\circle{15}}
 \put(80.5,30){\circle{15}}
 \put(98.5,48){\circle{15}}
 \put(57.5,43){\line(-1,-1){5.7}}
 \put(67.5,43){\line(1,-1){7.8}}
 \put(93.5,43){\line(-1,-1){7.8}}
 \put(103.5,43){\line(1,-1){5.7}}
 \put(57.5,43){\circle*{1.2}}
 \put(67.5,43){\circle*{1.2}}
 \put(93.5,43){\circle*{1.2}}
 \put(103.5,43){\circle*{1.2}}
 \put(75.5,35.2){\circle*{1.2}}
 \put(85.5,35.2){\circle*{1.2}}
 \put(70.5,42){\mbox{$I_{n-1}$}}
 \put(86,42){\mbox{$I_n$}}
 \put(59.5,35){\mbox{$\mathbb{S}^2_{n-1}$}}
 \put(79.5,17){\mbox{$\mathbb{S}^2_n$}}
 \put(95.5,35){\mbox{$\mathbb{S}^2_{n+1}$}}
 \end{picture}
 \caption{A loose zigzag necklace}
 \end{figure}
such that their geodesic distance is $\pi a/2$. The segment
$I_n$ now connects the points $0^{(n)}\equiv p_1^{(n)} \in
\mathbb{S}^2_n$ and $d^{(n)}\equiv p_2^{(n+1)} \in
\mathbb{S}^2_{n+1}$. \\ [.5em]
While the diagonal part (\ref{ren sphere G}) of the matrix $Q_0$
does not depend on the way we arrange the spheres, the
off-diagonal parts differ and now become
 \begin{eqnarray}
 Q_0^{i,i\pm 1} &\!=\!& -\, {1\over 8\sqrt\pi}\, {\Gamma\left(
 {1\over 4}+{t\over 2} \right) \over \Gamma\left(
 {3\over 4}+{t\over 2} \right)}\, {1\over \cos \pi\left(
 {1\over 4}+{t\over 2} \right)}\,, \\ 
 Q_0^{i,i\pm 2} &\!=\!& Q_0^{21}(k^2) = -\, {1\over 4\cos(\pi t)}\,,
 \end{eqnarray}
for the zigzag and straight case, respectively, with the notation
we have adopted. In the same way, the dispersion relation
(\ref{spectral5}) becomes
 \begin{equation} \label{loose}
 Q_0^{11} Q_0^{11} - Q_0^{1j} Q_0^{1j} -2\alpha^2k\, \cot(kd)\,
 Q_0^{11} -2\alpha^2k\, {Q_0^{1j}\over\sin(kd)}\,\cos\theta - \alpha^4
 k^2= 0
 \end{equation}
with $j=2,3$ in Examples II and I, respectively. Let us remark
that the condition with $j=2$ is valid whenever all the $Q_0^{12}$
are the same. Hence the spectrum does not change when we rotate an
arbitrary semi-infinite part of the chain around the axis given by
the appropriate connecting segment, such that, geometrically speaking,
the zigzag chain need not be periodic.


\subsection{Tight necklaces}

In a similar way, one can treat periodic sphere chains which are
connected through points where they touch (i.e. shrinking the line
segments to zero), with the boundary conditions (\ref{minbc-per}) replaced
by (\ref{minbc point}) at each junction. We shall consider again
two particular situations analogous to the periodic chains
discussed above:
\\ [.5em]
{\sl Example III:} Suppose that the junctions are situated at
antipodal points as sketched in Fig.~3,
 \begin{figure}
 \setlength\unitlength{1mm}
 \begin{picture}(65,45)(10,15)
 \thicklines
 \put(66.5,30){\circle{15}}
 \put(80.5,30){\circle{15}}
 \put(94.5,30){\circle{15}}
 \put(59.5,30){\circle*{1.2}}
 \put(73.5,30){\circle*{1.2}}
 \put(87.5,30){\circle*{1.2}}
 \put(101.5,30){\circle*{1.2}}
 \put(63.5,17){\mbox{$\mathbb{S}^2_{n-1}$}}
 \put(79.5,17){\mbox{$\mathbb{S}^2_n$}}
 \put(91.5,17){\mbox{$\mathbb{S}^2_{n+1}$}}
 \put(53,30){\mbox{$\dots$}}
 \put(103,30){\mbox{$\dots$}}
 \end{picture}
 \caption{A tight straight necklace}
 \end{figure}
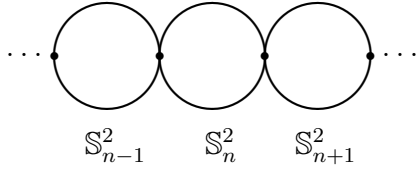
being obtained by identifying the points $p_1^{(n)} \in
\mathbb{S}^2_n$ and $p_3^{(n+1)} \in \mathbb{S}^2_{n+1}$.
 \\ [.5em]
\noindent {\sl Example IV:} The tight zigzag chain in Fig.~4 is
 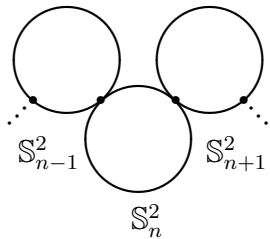
\begin{figure}
 \setlength\unitlength{1mm}
 \begin{picture}(75,45)(10,15)
 \thicklines
 \put(71,40.5){\circle{15}}
 \put(80.5,30){\circle{15}}
 \put(90,40.5){\circle{15}}
 \put(66.5,35.2){\circle*{1.2}}
 \put(94.5,35.2){\circle*{1.2}}
 \put(75.5,35.2){\circle*{1.2}}
 \put(85.5,35.2){\circle*{1.2}}
 \put(64.5,27){\mbox{$\mathbb{S}^2_{n-1}$}}
 \put(79.5,18){\mbox{$\mathbb{S}^2_n$}}
 \put(89.5,27){\mbox{$\mathbb{S}^2_{n+1}$}}
 \put(64.8,33){\mbox{$\cdot$}}
 \put(63.8,32){\mbox{$\cdot$}}
 \put(62.8,31){\mbox{$\cdot$}}
 \put(94.9,33){\mbox{$\cdot$}}
 \put(95.9,32){\mbox{$\cdot$}}
 \put(96.9,31){\mbox{$\cdot$}}
 \end{picture}
 \caption{A tight zigzag necklace}
 \end{figure}
obtained by identifying the points $p_1^{(n)} \in \mathbb{S}^2_n$
and $p_2^{(n+1)} \in \mathbb{S}^2_{n+1}$. \\ [.5em]
The dispersion relation now reads
 \begin{equation} \label{tight}
 Q_0^{11} Q_0^{11} - Q_0^{1j} Q_0^{1j} -2\alpha\, Q_0^{1j}
 \,\cos\theta + \alpha^2= 0
 \end{equation}
with $j=2,3$ corresponding to the Examples IV and III,
respectively.


\setcounter{equation}{0}
\section{Square bead carpets}

So far we have considered only ``manifolds'' with a linear
structure. Having in mind essential differences between spectra of
periodic Schr\"odinger operators in different dimensions
to detect, it is
useful to also to look systems which are periodic in more
than one direction; we will do this again by first analyzing simple
examples. This time we arrange our spherical ``beads'' into a
square lattice, coupling them either by line segments or directly
through touching points. \\ [.5em]
{\sl Example V:} Suppose that the connecting segments are attached
at four equally spaced points at the sphere equator as sketched in
Fig.~5,
 \begin{figure}
 \setlength\unitlength{1mm}
 \begin{picture}(125,45)(10,15)
 \thicklines
 \put(55.5,80){\circle{15}}
 \put(80.5,80){\circle{15}}
 \put(105.5,80){\circle{15}}
 \put(40.5,80){\line(1,0){8}}
 \put(62.5,80){\line(1,0){11}}
 \put(87.5,80){\line(1,0){11}}
 \put(112.5,80){\line(1,0){8}}
 \put(48.5,80){\circle*{1.2}}
 \put(62.5,80){\circle*{1.2}}
 \put(73.5,80){\circle*{1.2}}
 \put(87.5,80){\circle*{1.2}}
 \put(98.5,80){\circle*{1.2}}
 \put(112.5,80){\circle*{1.2}}
 \put(55.5,55){\circle{15}}
 \put(80.5,55){\circle{15}}
 \put(105.5,55){\circle{15}}
 \put(40.5,55){\line(1,0){8}}
 \put(62.5,55){\line(1,0){11}}
 \put(87.5,55){\line(1,0){11}}
 \put(112.5,55){\line(1,0){8}}
 \put(48.5,55){\circle*{1.2}}
 \put(62.5,55){\circle*{1.2}}
 \put(73.5,55){\circle*{1.2}}
 \put(87.5,55){\circle*{1.2}}
 \put(98.5,55){\circle*{1.2}}
 \put(112.5,55){\circle*{1.2}}
 \put(55.5,30){\circle{15}}
 \put(80.5,30){\circle{15}}
 \put(105.5,30){\circle{15}}
 \put(40.5,30){\line(1,0){8}}
 \put(62.5,30){\line(1,0){11}}
 \put(87.5,30){\line(1,0){11}}
 \put(112.5,30){\line(1,0){8}}
 \put(48.5,30){\circle*{1.2}}
 \put(62.5,30){\circle*{1.2}}
 \put(73.5,30){\circle*{1.2}}
 \put(87.5,30){\circle*{1.2}}
 \put(98.5,30){\circle*{1.2}}
 \put(112.5,30){\circle*{1.2}}
 \put(55.5,87){\line(0,1){8}}
 \put(80.5,87){\line(0,1){8}}
 \put(105.5,87){\line(0,1){8}}
 \put(55.5,15){\line(0,1){8}}
 \put(80.5,15){\line(0,1){8}}
 \put(105.5,15){\line(0,1){8}}
 \put(55.5,62){\line(0,1){11}}
 \put(80.5,62){\line(0,1){11}}
 \put(105.5,62){\line(0,1){11}}
 \put(55.5,37){\line(0,1){11}}
 \put(80.5,37){\line(0,1){11}}
 \put(105.5,37){\line(0,1){11}}
 \put(55.5,87){\circle*{1.2}}
 \put(80.5,87){\circle*{1.2}}
 \put(105.5,87){\circle*{1.2}}
 \put(55.5,62){\circle*{1.2}}
 \put(80.5,62){\circle*{1.2}}
 \put(105.5,62){\circle*{1.2}}
 \put(55.5,37){\circle*{1.2}}
 \put(80.5,37){\circle*{1.2}}
 \put(105.5,37){\circle*{1.2}}
 \put(55.5,73){\circle*{1.2}}
 \put(80.5,73){\circle*{1.2}}
 \put(105.5,73){\circle*{1.2}}
 \put(55.5,48){\circle*{1.2}}
 \put(80.5,48){\circle*{1.2}}
 \put(105.5,48){\circle*{1.2}}
 \put(55.5,23){\circle*{1.2}}
 \put(80.5,23){\circle*{1.2}}
 \put(105.5,23){\circle*{1.2}}
 \put(62.5,59){\mbox{\scriptsize{$I_{m-{1\over2},n}$}}}
 \put(87.5,59){\mbox{\scriptsize{$I_{m+{1\over2},n}$}}}
 \put(81.5,68){\mbox{\scriptsize{$I_{m,n+{1\over2}}$}}}
 \put(81.5,39){\mbox{\scriptsize{$I_{m,n-{1\over2}}$}}}
 \put(79,54){\mbox{$\mathbb{S}^2_n$}}
 \put(87,50){\mbox{\scriptsize{$p^{(n)}_1\equiv 0$}}}
 \put(68.5,63){\mbox{\scriptsize{$p^{(n)}_2\equiv 0$}}}
 \put(63,50){\mbox{\scriptsize{$p^{(n)}_3\equiv d$}}}
 \put(81.5,44.5){\mbox{\scriptsize{$p^{(n)}_4\equiv d$}}}
 \end{picture}
 \caption{A loose square bead carpet}
 \end{figure}
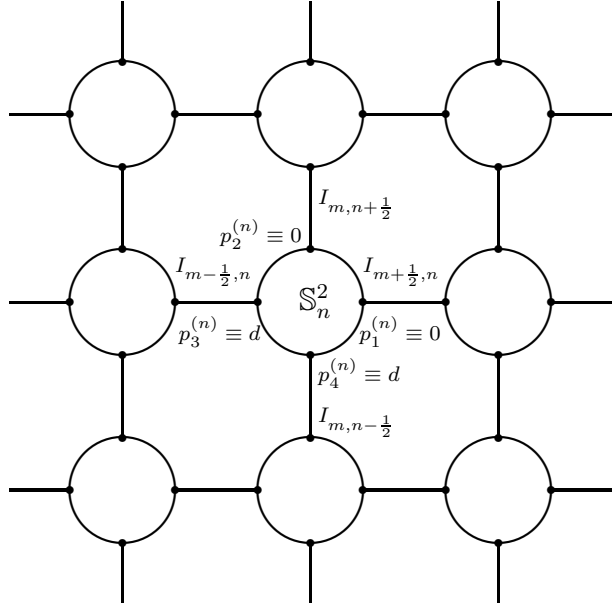
where the labeling of the junctions and segments is indicated.
With the notation introduced in Fig.5 the boundary conditions
defining the Hamiltonian read
 \begin{eqnarray*}
 b\left(f_0^{(n,m)}, p_1^{(n,m)}\right) &\!=\!& -\alpha \left(
 f_1^{(n+{1\over2},m)} \right)' \left( 0^{(n+{1\over2},m)} \right)
 \,, \\ f_1^{(n+{1\over2},m)} \left( 0^{(n+{1\over2},m)}
 \right)&\!=\!& \bar\alpha a\left(f_0^{(n,m)}, p_1^{(n,m)}\right)
 \,, \\ b\left(f_0^{(n,m)}, p_3^{(n,m)}\right) &\!=\!& -\alpha \left(
 f_1^{(n-{1\over2},m)} \right)' \left( d^{(n-{1\over2},m)} \right)
 \,, \\ f_1^{(n-{1\over2},m)} \left( d^{(n-{1\over2},m)}
 \right)&\!=\!& \bar\alpha a\left(f_0^{(n,m)}, p_3^{(n,m)}\right)
 \,, \\ b\left(f_0^{(n,m)}, p_2^{(n,m)}\right) &\!=\!& -\alpha \left(
 f_1^{(n,m+{1\over2})} \right)' \left( 0^{(n,m+{1\over2})} \right)
 \,, \\ f_1^{(n,m+{1\over2})} \left( 0^{(n,m+{1\over2})}
 \right)&\!=\!& \bar\alpha a\left(f_0^{(n,m)}, p_2^{(n,m)}\right)
 \,, \\ b\left(f_0^{(n,m)}, p_4^{(n,m)}\right) &\!=\!& -\alpha \left(
 f_1^{(n,m-{1\over2})} \right)' \left( 0^{(n,m-{1\over2})} \right)
 \,, \\ f_1^{(n,m-{1\over2})} \left( 0^{(n,m-{1\over2})}
 \right)&\!=\!& \bar\alpha a\left(f_0^{(n,m)},
 p_4^{(n,m)}\right)\,.
 \end{eqnarray*}
The dispersion relation is derived as in the previous section, but
it becomes rather cumbersome. It is useful to introduce the following
notation:
 \begin{eqnarray*}
 \Delta &\!:=\!& {1\over k^2}\, \left(Q_0^{12} Q_0^{12} - Q_0^{11}
 Q_0^{11} \right) + {2\alpha^2\over k\sin(kd)}\, \left( Q_0^{11}
 \cos(kd) + Q_0^{12} \right) + \alpha^4\,, \\
 a_j &\!:=\!& Q_0^{1,j+1} \Delta + \left( {Q_0^{1,j+1}\over k^2}
 -(-1)^j {\alpha^2\over k\sin(kd)} \right) \left(Q_0^{12} Q_0^{12}
 + Q_0^{13} Q_0^{13} \right) \\ && +\, 2\, Q_0^{12} Q_0^{13} \left(
 {Q_0^{1,2-j}\over k^2} +(-1)^j {\alpha^2\over k\sin(kd)}
 \right)\,,\; j=0,1\,, \\
 b_0 &\!:=\!& {1\over k^2\sin^2(kd)} \left\lbrack {{Q_0^{11} Q_0^{11}
 - Q_0^{12} Q_0^{12}}\over k\sin(kd)}\, \cos(kd) + \alpha^2 Q_0^{11}
 \right\rbrack - {\Delta\cos(kd)\over k\sin(kd)}\,, \\
 b_1 &\!:=\!& {1\over k^2\sin^2(kd)} \left\lbrack {{Q_0^{11} Q_0^{11}
 - Q_0^{12} Q_0^{12}}\over k\sin(kd)} - \alpha^2 Q_0^{11} \right\rbrack
 \,,\\
 c_j &\!:=\!& {\alpha\over k\sin(kd)} \Bigg\lbrack \alpha^2
 Q_0^{1,4-j} + {Q_0^{1,4-j} Q_0^{11} \cos(kd)\over k\sin(kd)} \\ &&
 \phantom{AAAAAA} -\,
 {{Q_0^{1,j+1}Q_0^{11} + Q_0^{1,4-j} Q_0^{12} \cos(kd)}\over
 k\sin(kd)} \Bigg\rbrack\,, \; j=1,2\,.
 \end{eqnarray*}
Using this notation, we can write the spectral condition as
 \begin{eqnarray} \label{loose carpet}
 \lefteqn{ (a_0^2-a_1^2) (b_0^2-b_1^2) + (c_1^2-c_2^2)^2 -
 2\big\lbrack (c_1+c_2)^2 (a_0b_0+a_1b_1) } \nonumber \\ &&
 - 2c_1c_2 (a_0+a_1) (b_0+b_1) \big\rbrack \nonumber
 \\ && + 2\alpha\Delta \left\lbrack (a_0b_0+a_1b_1-c_1^2)c_1 +
 a_0b_1+a_1b_0 - c_1c_2)c_2 \right\rbrack (\cos\theta_1 +
 \cos\theta_2) \nonumber \\ && + 2\alpha^2\Delta^2 \left\lbrack (c_1^2-c_2^2)
 \cos(\theta_1+\theta_2) + (c_1^2-a_1b_1) \cos(\theta_1-\theta_2)
 + c_1^2 -a_0b_0 \right\rbrack \nonumber \\ && - 2\alpha^3\Delta^3 c_1
 (\cos\theta_1 + \cos\theta_2) + \alpha^4\Delta^4 = 0\,.\,
 \end{eqnarray}
where $\theta_1, \theta_2$ are the quasimomentum components.
 \\ [.5em]

\noindent {\sl Example VI:} This arises from Example V by shrinking the connecting
 \begin{figure}
 \setlength\unitlength{1mm}
 \begin{picture}(125,45)(10,15)
 \thicklines
 \put(66.5,58.4){\circle{15}}
 \put(80.5,58.4){\circle{15}}
 \put(94.5,58.4){\circle{15}}
 \put(59.5,58.4){\circle*{1.2}}
 \put(73.5,58.4){\circle*{1.2}}
 \put(87.5,58.4){\circle*{1.2}}
 \put(101.5,58.4){\circle*{1.2}}
 \put(77,57.4){\mbox{\scriptsize{$\mathbb{S}^2_{m,n+1}$}}}
 \put(53,58.4){\mbox{$\dots$}}
 \put(103,58.4){\mbox{$\dots$}}
 \put(66.5,44.2){\circle{15}}
 \put(80.5,44.2){\circle{15}}
 \put(94.5,44.2){\circle{15}}
 \put(59.5,44.2){\circle*{1.2}}
 \put(73.5,44.2){\circle*{1.2}}
 \put(87.5,44.2){\circle*{1.2}}
 \put(101.5,44.2){\circle*{1.2}}
 \put(62.5,43.2){\mbox{\scriptsize{$\mathbb{S}^2_{m-1,n}$}}}
 \put(78,43.2){\mbox{\scriptsize{$\mathbb{S}^2_{m,n}$}}}
 \put(90.5,43.2){\mbox{\scriptsize{$\mathbb{S}^2_{m+1,n}$}}}
 \put(53,44.2){\mbox{$\dots$}}
 \put(103,44.2){\mbox{$\dots$}}
 \put(66.5,30){\circle{15}}
 \put(80.5,30){\circle{15}}
 \put(94.5,30){\circle{15}}
 \put(59.5,30){\circle*{1.2}}
 \put(73.5,30){\circle*{1.2}}
 \put(87.5,30){\circle*{1.2}}
 \put(101.5,30){\circle*{1.2}}
 \put(77,29){\mbox{\scriptsize{$\mathbb{S}^2_{m,n-1}$}}}
 \put(53,30){\mbox{$\dots$}}
 \put(103,30){\mbox{$\dots$}}
 \put(66.5,65.5){\circle*{1.2}}
 \put(80.5,65.5){\circle*{1.2}}
 \put(94.5,65.5){\circle*{1.2}}
 \put(66.5,51.3){\circle*{1.2}}
 \put(80.5,51.3){\circle*{1.2}}
 \put(94.5,51.3){\circle*{1.2}}
 \put(66.5,37.1){\circle*{1.2}}
 \put(80.5,37.1){\circle*{1.2}}
 \put(94.5,37.1){\circle*{1.2}}
 \put(66.5,22.9){\circle*{1.2}}
 \put(80.5,22.9){\circle*{1.2}}
 \put(94.5,22.9){\circle*{1.2}}
 \put(66,18){\vdots}
 \put(80,18){\vdots}
 \put(94,18){\vdots}
 \put(66,67.2){\vdots}
 \put(80,67.2){\vdots}
 \put(94,67.2){\vdots}
 \end{picture}
 \caption{A tight square bead carpet}
 \end{figure}
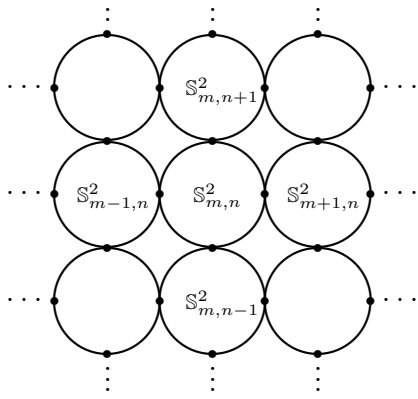
seg\-ments to zero, as indicated in Fig.~6 where the labeling of 
the junctions is the same as in the previous example. After a 
straightforward calculation we find that the spectral condition 
now takes the form 
 \begin{eqnarray} \label{tight carpet}
 \lefteqn{ (Q_0^{11} Q_0^{11} - Q_0^{13} Q_0^{13})^2 -4Q_0^{12}
 Q_0^{12}(Q_0^{11} - Q_0^{13})^2 + 2\alpha  \big\lbrack Q_0^{13}
 Q_0^{13} Q_0^{13} } \nonumber \\ && - Q_0^{11}
 Q_0^{11} Q_0^{13} +2 Q_0^{11} Q_0^{12} Q_0^{12} - 2 Q_0^{12}
 Q_0^{12} Q_0^{13} \big\rbrack (\cos\theta_1 + \cos\theta_2) \nonumber \\
 && +2\alpha^2 \left\lbrack Q_0^{13} Q_0^{13} - Q_0^{11} Q_0^{11}
 +2(Q_0^{13} Q_0^{13} - Q_0^{12} Q_0^{12} ) \right\rbrack
 \cos\theta_1 \cos\theta_2 \nonumber \\ && -2\alpha^3 Q_0^{13} (\cos\theta_1 +
 \cos\theta_2) + \alpha^4 = 0\,.
 \end{eqnarray}


\setcounter{equation}{0}
\section{Gap dominance at large energies}

As customary in periodic systems the spectrum of the above described
operators (which we denote by $H_\mathrm{I},\dots,H_\mathrm{VI}$
according to the example number) has band structure. To see how
the gap width and the band width are related at high energies,
consider first the points
 \begin{equation} \label{pole point}
 k'_n := {\pi n\over d}\,,\quad k''_n:= {\sqrt{n(n\!+\!1)}\over a}\,,
 \quad n=1,2,\dots\,,
 \end{equation}
for which $\sin(dk'_n) = \cos(dk'_n) = \cos \pi\left( {1\over 4} +
{1\over 2}t(k''_{2n}) \right) =0$, such that the functions
$Q_0^{ij}$ and $Q_1^{ij}$ have poles. Thus it is natural to look for
spectral bands in the vicinity of these points. We fix
$\epsilon>0$ and denote by $J'_n= [k'_n-\delta'_n,k'_n+\delta'_n]$
the maximal closed neighbourhood of the point $k'_n$ in which the
inequality
 $$ 
 |\sin(kd)| \le k^{-\epsilon}
 $$ 
is satisfied; in the same way the intervals $J''_n=
[k''_n-\delta''_n,k''_n+\tilde\delta''_n]$ and $J'''_n=
[k''_{2n}-\delta'''_n,k'_{2n}+\tilde\delta'''_n]$ correspond to
the inequalities
 \begin{equation} \label{loose est}
 |\cos(kd)| \le k^{-\epsilon} \quad\mathrm{and}\quad \left|\cos \pi
 \left( {1\over 4} + {t\over 2}\right)\right| \le k^{-\epsilon}\,,
 \end{equation}
respectively. It is clear that all the $\delta'_n,\dots,
\tilde\delta'''_n$ are strictly positive, and it is not difficult
to check that
 $$ 
 \delta'_n \sim d^{-1} (k'_n)^{-\epsilon}\,,\quad \delta''_n,
 \tilde\delta''_n \sim 2(\pi a)^{-1} (k'_n)^{-\epsilon}\,,\quad
 \delta'''_n,
 \tilde\delta'''_n \sim 4(\pi a)^{-1} (k'_{2n})^{-\epsilon}
 $$ 
as $n\to \infty$. Our aim is to show that for a sufficiently high
energy the spectral gaps contain the complement of the above
intervals. More specifically, define
 $$ 
 \Omega_K := [K,\infty) \setminus \bigcup_{n=1}^{\infty} (J'_n
 \cup J''_n \cup J'''_n)
 $$ 
for a fixed $K>0$.In this set, we have $(\sin(kd))^{-1} =
\OO(k^\epsilon)$ as $k\to\infty$, and similarly
 $$ 
 Q_0^{11} = \OO(k^\epsilon)\,, \quad  Q_0^{12} = \OO(k^{\epsilon-1})\,,
 \quad Q_0^{13} = \OO(k^\epsilon)\,,
 $$ 
where the first relation was derived using the asymptotic
relation
 $$ 
 {\Gamma\left(
 {1\over 4}+{t\over 2} \right) \over \Gamma\left(
 {3\over 4}+{t\over 2} \right)}\,= {2\over t} \left( 1+\OO(t^{-2})
 \right)\,,
 $$ 
which follows from the Stirling formula. These relations show that
the left-hand-side of (\ref{loose}) behaves in $\Omega_K$ as
 $$ 
 -\alpha^4 k^2 +\OO(k^{1+2\epsilon})
 $$ 
for $k\to\infty$, and therefore it diverges uniformly in $\theta$
as long as $0<\epsilon< {1\over 2}$. Consequently, there is $K>0$
such that
 $$ 
 {\rm spec} H_\mathrm{I} \cap \Omega_K = {\rm spec} H_\mathrm{I} \cap \Omega_K
 =\emptyset\,.
 $$ 

Let us pass to the relation (\ref{loose carpet}). Notice first
that $\Delta\to \alpha^4$ as $k\to\infty$ in $\Omega_K$.
Furthermore, for $0<\epsilon< {1\over 2}$ we have
 $$ 
 a_0 = \OO(k^\epsilon)\,, \;  a_1 = \OO(k^{2\epsilon-1})\,,
 \; b_j = \OO(k^{\epsilon-1})\,, \;  c_1 =
 \OO(k^{2\epsilon-1})\,, \; c_2 = \OO(k^{4\epsilon-1})\,.
 $$ 
Consequently for $\epsilon<{1\over 4}$,
the left-hand side of the spectral condition tends
to $\alpha^8\ne 0$, which implies
 $$ 
 {\rm spec} (H_\mathrm{V}) \cap \Omega_K =\emptyset
 $$ 
for $K$ large enough.

The tight necklaces and carpets exhibit a different behaviour. Now we
replace the intervals $J''_n,\, J'''_n$ defined by
(\ref{loose est}) by $\hat J''_n
=[k''_n-\eta''_n, k''_n-\tilde\eta''_n]$ and $\hat
J'''_n=[k''_{2n}-\eta'''_n, k_{2n}-\tilde\eta'''_n]$ given in a
similar way by
 \begin{equation} \label{tight est}
 |\cos(kd)| \le (\ln k)^{-\epsilon} \quad\mathrm{and}\quad \left|\cos \pi
 \left( {1\over 4} + {t\over 2}\right)\right| \le (\ln
 k)^{-\epsilon}\,.
 \end{equation}
It is straightforward to check that
 $$ 
 \eta''_n, \tilde\eta''_n \sim 2(\pi a)^{-1} (\ln k'_n)^{-\epsilon}
 \,,\quad \eta'''_n, \tilde\eta'''_n \sim 4(\pi a)^{-1}
 (\ln k'_{2n})^{-\epsilon}\,.
 $$ 
Consider the set $\hat\Omega_K := [K,\infty) \setminus
\bigcup_{n=1}^{\infty} (\hat J''_n \cup \hat J'''_n)$ with a fixed
$K>1$. If $k\to\infty$ in this set, the following estimates
hold:
 $$ 
 Q_0^{11} = A\,\ln k+\OO((\ln k)^\epsilon)\,, \quad
 Q_0^{12} = \OO(k^{-1}(\ln k)^{\epsilon})\,,
 \quad Q_0^{13} = \OO((\ln k)^\epsilon)\,,
 $$ 
with $A\ne 0$. These relations show that the left-hand side of
(\ref{tight}) diverges for $\epsilon<1$ like $(\ln k)^2$, uniformly
in $\theta$ as $k\to \infty$ within $\hat\Omega_K$. By the same
token,
the left-hand side of (\ref{tight carpet}) diverges for
$\epsilon<1$ like $(\ln k)^4$, uniformly in $\theta_1,\theta_2$. We
infer that there is a $K>1$ such that
 $$ 
 {\rm spec}H_\mathrm{III} \cap \hat\Omega_K = {\rm spec}H_\mathrm{IV} \cap
 \hat\Omega_K = {\rm spec}H_\mathrm{VI} \cap \hat\Omega_K =\emptyset\,.
 $$ 

Now it is easy to estimate the band and gap widths. The points
$E'_n = (k'_n)^2$ and $E''_n = (k''_n)^2$ around which the bands
concentrate are asymptotically like $c' n^2$ and $c" n^2$,
respectively, by (\ref{pole point}).
The widths of the excluded intervals behave, in the case of a loose
connection, as
 $$ 
 |J'_n|\,,\; |J'_n|\,,\; |J'_n|\sim
 \mathrm{const}\,n^{1-\epsilon}\,.
 $$ 
Hence the total length $B_n$ of the bands contained in the union
of the intervals $J'_n,\, J''_n$, and $J'''_n$ is of order $B_n
\aleq \mathrm{const}\,n^{1-\epsilon}$, and the total length $L_n$
of the adjacent gaps is $L_n \ageq \mathrm{const}\,
(n\!-\!n^{1-\epsilon}) \approx \mathrm{const}\, n$. In the case of
a tight connection the band length is estimated instead by $B_n
\aleq \mathrm{const}\,n(\ln n)^{-\epsilon}$ which still gives gap
length increasing linearly with $n$. We sum up our discussion with
the following result:
 \begin{proposition} \label{final}
 For loosely connected necklaces and carpets the band-to-gap ratio
 satisfies the bound
 $$ 
 {B_n\over L_n} \aleq \mathrm{const}\,n^{-\epsilon}
 $$ 
 as $n\to\infty$, with a positive $\epsilon<{1\over 2}$ in Examples I and
 II, and $\epsilon<{1\over 4}$ in Example~V. On the other hand, for
 the tightly connected necklaces and carpets in Examples III, IV, and
 VI, we have
 $$ 
 {B_n\over L_n} \aleq \mathrm{const}\,(\ln n)^{-\epsilon}
 $$ 
 as $n\to\infty$, with any positive $\epsilon<1$.
 \end{proposition}


\subsection*{Acknowledgment}

The research has been partially supported by SFB (project $\#$ 288),
GA AS (contract $\#$ 1048101), DFG (Grant $\#$ 436 RUS 113/572/1),
INTAS (Grant $\#$ 00-257), and RFBR (Grant $\#$ 02-01-00804).
.

\end{document}